\pdfoutput=1
\documentclass[11pt]{article}

\usepackage[T1]{fontenc}
\usepackage[utf8]{inputenc}
\usepackage{lmodern}
\usepackage[english]{babel}
\usepackage[margin=1in]{geometry}
\usepackage{amsmath,amssymb}
\usepackage{graphicx}
\usepackage{booktabs}
\usepackage{caption}
\usepackage{subcaption}
\usepackage[round,authoryear]{natbib}
\usepackage[hidelinks]{hyperref}

\DeclareMathOperator{\clip}{clip}
\title{\bfseries Reconstructing the Human Development Index at the
Individual Level from Survey Microdata: Development and Validation of a
Methodology}

\author{%
  Tcharlies Bachmann Schmitz\thanks{Corresponding author:
  \texttt{tcharlies.schmitz@px.center}} \\
  \textit{PX Center, Brazil}
  \and
  Natalia Veronez da Cunha\thanks{\texttt{nat\_cunha@hotmail.com}} \\
  \textit{Universidade do Planalto Catarinense (UNIPLAC), Lages, SC, Brazil}
  \and
  Isabelle Ferreira\thanks{\texttt{isabelle.ferreira@uniplaclages.edu.br}} \\
  \textit{Universidade do Planalto Catarinense (UNIPLAC), Lages, SC, Brazil}
}

\date{}

\begin{document}
\maketitle

\begin{abstract}
\noindent This study proposes and validates a methodology for reconstructing
the Human Development Index (HDI) from individual-level survey microdata,
enabling disaggregated estimates for demographic subpopulations while
maintaining coherence with the official HDI. Data from 40,185 respondents in
21 countries from the European Social Survey (ESS7, 2014) were analysed,
combined with World Bank indicators for life expectancy, gross national income
per capita, and expected years of schooling. The health dimension was estimated
from national life expectancy, adjusted by self-rated health; income, by a
lognormal multiplier calibrated by the Gini coefficient; and education, by
combining observed and expected years of schooling. The dimensions were
aggregated by geometric mean, weighted by sampling weights, with confidence
intervals estimated via the delta method. Among the 39,706 individuals with
valid data (98.8\%), the reconstructed HDI showed high agreement with the
official 2014 HDI, with a mean absolute error of 0.0099 and a Pearson
correlation of 0.917. Consistent differences across subgroups evidenced the
methodology's convergent validity and its capacity to produce disaggregated
estimates of human development, expanding the HDI's potential for analyses of
socioeconomic and demographic inequalities.

\vspace{1em}
\end{abstract}

\section{Introduction}

Since its introduction in the human development reports of the United Nations
Development Programme (UNDP), the Human Development Index (HDI) has established
itself as the most widespread synthesis of the notion that the development of a
society cannot be reduced to its income, but also encompasses the capacity to
live a long and healthy life and to access knowledge. The methodological
formalisation of the index is largely due to \citet{anand1994}, who established
the three-dimensional structure --- health, education and income --- aggregated
into a single value, subsequently consolidated in the technical notes of the
official reports \citep{undp2015}. The index is, by construction, an aggregate
measure: each country receives a value computed from national accounts and
vital statistics, and no individual, taken in isolation, has an HDI.

This aggregate nature, however, imposes a relevant limitation when the
analytical interest falls on subpopulations. Public policy and social
diagnostics frequently require knowing how human development is distributed
within a country --- between men and women, across age cohorts, across
socioeconomic strata. Estimating the HDI for such breakdowns necessarily
requires descending to the individual level, since only microdata allow the
population to be regrouped according to criteria that do not coincide with
national borders. Hence a fundamental methodological tension arises: the index
was conceived for the aggregate level, but its application to subgroups
requires an individual formulation that, at the same time, remains coherent
with the official value when reaggregated.

Constructing an individual-level measure demands care with a well-known risk in
quantitative social science: the ecological fallacy, seminally described by
\citet{robinson1950}, according to which correlations observed between group
means cannot be transferred, without further justification, to the behaviour of
the individuals who compose those groups. This risk is especially pertinent to
the health dimension, whose life expectancy is, by nature, a population-level
quantity.

The present work therefore adopts the design principle of anchoring each
dimension in sources appropriate to its level of measurement: official national
life expectancy serves as an anchor, and individual variation enters only as
within-country modulation. The central objective of this study is thus to
propose and validate a statistically robust methodology for computing the HDI
at the individual level, applied to the microdata of the European Social Survey,
Round 7 \citep{ess7}, and validated against the official UNDP HDI for the year
2014 \citep{undp2015}.

Two research questions guide the investigation. First, is it possible to
reconstruct, from individual microdata of an international survey, a
country-level HDI that reproduces the official index computed from entirely
independent sources? Second, does the individual-level methodology allow
internally consistent demographic disaggregations, with honest quantification of
uncertainty?

This work makes two contributions. The first is methodological: we present a
formulation of the HDI at the individual level that preserves the normalisation
functions and the aggregation rule of the official index, anchors the health
dimension in national values, derives uncertainty from the actual sampling
variance, and treats income through a single route coherent with the UNDP
standard. The second is empirical: we demonstrate that the HDI reconstructed
from microdata approaches the official HDI with a mean absolute error below
0.01, offering evidence of convergent validity. The remainder of the article is
organised as follows. Section~\ref{sec:background} reviews the pertinent
literature. Section~\ref{sec:methods} describes the data sources and the
individual-level methodology, with the equations that define it.
Section~\ref{sec:results} presents the validation results and the demographic
breakdowns. Section~\ref{sec:discussion} interprets the findings and discusses
limitations. Section~\ref{sec:conclusion} concludes and points to future work.

\section{Background and Related Work}
\label{sec:background}

\subsection{The construction of the Human Development Index}

The official HDI results from the geometric mean of three dimensional indices
normalised between zero and one. The health dimension is represented by life
expectancy at birth; education, by the combination of mean years of schooling of
the adult population and expected years of schooling of school-age children; and
income, by the logarithm of gross national income per capita at purchasing power
parity. The choice of the geometric mean, replacing the arithmetic mean adopted
until 2009, reflects a deliberate decision to penalise unequal performance
across dimensions, so that a pronounced deficiency in any of them cannot be
fully compensated by excellence in another \citep{anand1994,undp2015}. The
logarithmic normalisation of income, in turn, materialises the hypothesis of
decreasing marginal utility of income, according to which successive increments
of income contribute ever less to the expansion of human capabilities. These
foundations are essential to the present work, since the individual-level
reconstruction proposed here fully preserves the normalisation functions and the
aggregation rule of the official index, altering only the level at which the
inputs are measured.

\subsection{The problem of ecological inference}

Transposing statistical relations estimated on aggregate data to the individual
level is one of the classic pitfalls of quantitative social research.
\citet{robinson1950} demonstrated that ecological correlations --- those
computed over group means --- may differ drastically, in magnitude and even in
sign, from the corresponding individual-level correlations. The implication for
HDI modelling is direct: estimating individual life expectancy from
characteristics observed only as aggregate means presupposes that the aggregate
relation holds at the individual level, a hypothesis that ecological inference
warns is unjustified. For this reason, the methodology adopted here does not
model life expectancy from aggregates; the official national value is treated as
a true anchor, and individual variation enters only as within-country
modulation, as detailed in Section~\ref{sec:methods}.

\subsection{Self-rated health and mortality}

The decision to modulate individual life expectancy by a self-rated health score
rests on extensive epidemiological evidence. The review by
\citet{idler1997}, covering twenty-seven community studies, established that the
subjective perception of one's own health status is an independent predictor of
mortality, even after controlling for objective indicators of morbidity and for
clinical assessments. This finding provides theoretical support for the
hypothesis that individuals who report better health, within the same country,
have above-average life expectancy, and vice versa. Self-rated health does not
replace a subgroup-specific life table, but it constitutes an empirically
legitimate modulator of life expectancy around the national anchor.

\subsection{Weighting in sample surveys}

Complex surveys such as the European Social Survey employ sampling designs that
combine stratification, clustering and unequal selection probabilities,
requiring the use of sampling weights so that population estimates are unbiased.
The sampling theory of \citet{kish1965} provides the framework for computing
weighted means and, above all, for the notion of effective sample size, which
adjusts the nominal size for the loss of efficiency introduced by the
variability of the weights. In the present work, all dimensional means and their
standard errors are computed with the ESS analytical weights, and the effective
sample size is reported as a measure of the actual precision of each estimate.

\section{Materials and Methods}
\label{sec:methods}

\subsection{Data sources}

The individual-level base is the European Social Survey, Round 7 \citep{ess7},
integrated file, edition 2.3, referring to the 2014 collection and covering
40,185 respondents distributed across 21 European countries, including Israel.
The round includes a rotating module on health inequalities, from which the
variables of interest are drawn. Each respondent is identified by country
(\texttt{cntry}), age (\texttt{agea}) and sex (\texttt{gndr}). Formal education
is measured in completed years of schooling (\texttt{eduyrs}). Household income
is recorded in national deciles (\texttt{hinctnta}, from 1 to 10).

Health is captured by the self-assessment of general health status
(\texttt{health}, from 1, very good, to 5, very bad) and by the presence of
hampering in daily activities due to illness or disability (\texttt{hlthhmp},
from 1 to 3). Lifestyle indicators --- fruit consumption (\texttt{etfruit}),
vegetable consumption (\texttt{eatveg}) and frequency of alcohol consumption
(\texttt{alcfreq}) --- are recoded monotonically. The estimates employ the ESS
analytical weight, defined as the product of the post-stratification weight and
the population weight (\texttt{anweight} = \texttt{pspwght} $\times$
\texttt{pweight}), appropriate for multi-country analyses.

The national inputs come from the World Development Indicators of the
\citet{worldbank2026}: life expectancy at birth (code
\texttt{SP.DYN.LE00.IN}), gross national income per capita at purchasing power
parity (code \texttt{NY.GNP.PCAP.PP.CD}) and expected years of schooling (code
\texttt{SE.SCH.LIFE}), all referring to 2014 or to the closest available year.
As a reproducibility note, we record that the microdata were obtained in an
automated way through the public GraphQL interface of Sikt/NSD, by means of the
asynchronous cycle of \texttt{startDownloadJob} and \texttt{pollDownloadJob}
requests, which returns a signed cloud-storage URL without requiring
authentication; the World Bank series are likewise open.

\subsection{Health dimension}

The health dimension is anchored in the official life expectancy of the country
and modulated by the relative position of each individual on a standardised
health score. Let $\mathit{LE}_{p}$ be the life expectancy at birth of country
$p$ and $z_{i}$ the health score of individual $i$, defined as the mean of the
standardised components of self-rated health (recoded so that higher values
indicate better health) and of absence of hampering. Individual life expectancy
is given by
\begin{equation}
\mathit{LE}_{i} = \mathit{LE}_{p(i)}
  + \beta\,\bigl( z_{i} - \bar{z}_{p(i)} \bigr),
\end{equation}
where $\bar{z}_{p}$ is the weighted mean of the score in country $p$ and $\beta$
is the gain, in years of life expectancy, associated with one standard deviation
of the health score; the conservative value $\beta = 2$ is adopted. Centring by
$\bar{z}_{p}$ is the decisive methodological decision: by subtracting the
national mean of the score, it is guaranteed that the weighted mean of
$\mathit{LE}_{i}$ within each country reproduces exactly the official life
expectancy $\mathit{LE}_{p}$, preserving the national anchor. Individual
variation enters only as internal modulation, without violating the ecological
inference discussed by \citet{robinson1950}, and the empirical justification for
the modulation rests on \citet{idler1997}. The health index results from the
UNDP normalisation,
\begin{equation}
I_{i}^{(h)} = \clip\!\left( \frac{\mathit{LE}_{i} - 20}{85 - 20},\; 0,\; 1
\right),
\end{equation}
where the bounds of 20 and 85 years are the minimum and maximum reference values
adopted by the UNDP and $\clip$ denotes truncation to the unit interval.

\subsection{Income dimension}

Individual income derives from the product between the country's gross national
income per capita at purchasing power parity, $\mathit{GNI}_{p}$, and a
multiplier that positions the individual according to their household income
decile. The multiplier derives from a lognormal income distribution with unit
mean over the population, calibrated by a Gini coefficient $G$. For a decile
$d \in \{1,\ldots,10\}$, the percentile position $p = (d - 0.5)/10$ and the
multiplier are defined as
\begin{equation}
m(d) = \exp\!\left( \sigma\,\Phi^{-1}(p) - \frac{\sigma^{2}}{2} \right),
\qquad
\sigma = \sqrt{2}\,\Phi^{-1}\!\left( \frac{G + 1}{2} \right),
\end{equation}
where $\Phi^{-1}$ is the quantile function of the standard normal distribution
and the relation between $\sigma$ and $G$ corresponds to the classic expression
for the Gini coefficient of a lognormal distribution,
$G = 2\,\Phi(\sigma/\sqrt{2}) - 1$. The value $G = 0.30$ is adopted as typical.
Individual income is $\mathit{income}_{i} = \mathit{GNI}_{p(i)} \cdot m(d_{i})$,
and the income index follows the UNDP logarithmic normalisation,
\begin{equation}
I_{i}^{(y)} = \clip\!\left(
  \frac{\ln \mathit{income}_{i} - \ln 100}{\ln 75000 - \ln 100},\; 0,\; 1
\right),
\end{equation}
with bounds of 100 and 75,000 international dollars. This route works directly
on gross national income per capita at purchasing power parity, exactly the base
used by the UNDP, which ensures conceptual coherence with the official index.

\subsection{Education dimension}

The education index combines, in equal parts, the individual's observed years of
schooling and the country's expected years of schooling, $\mathit{ESY}_{p}$,
each normalised by its reference ceiling,
\begin{equation}
I_{i}^{(e)} = \frac{1}{2}\left[
  \min\!\left( \frac{\mathit{eduyrs}_{i}}{15},\, 1 \right)
  + \min\!\left( \frac{\mathit{ESY}_{p(i)}}{18},\, 1 \right)
\right].
\end{equation}
Adopting a country-specific value of $\mathit{ESY}_{p}$ preserves the national
heterogeneity of this component.

\subsection{Aggregation by breakdown and uncertainty propagation}

For a breakdown $k$ (country, sex, age band or cross-classification), each
dimensional index is estimated by the mean weighted by the sampling weights,
\begin{equation}
\bar{I}_{k}^{(\cdot)} =
  \frac{\sum_{i \in k} w_{i}\, I_{i}^{(\cdot)}}{\sum_{i \in k} w_{i}},
\end{equation}
and its standard error derives from the actual weighted sampling variance,
\begin{equation}
\mathit{SE}_{k}^{(\cdot)} =
  \frac{\sqrt{\sum_{i \in k} w_{i}^{2}\,
    \bigl( I_{i}^{(\cdot)} - \bar{I}_{k}^{(\cdot)} \bigr)^{2}}}
       {\sum_{i \in k} w_{i}},
\end{equation}
which grounds uncertainty in the dispersion actually observed in the sample
\citep{kish1965}. The HDI of the breakdown is the geometric mean of the three
aggregated dimensions,
\begin{equation}
\mathit{HDI}_{k} = \bigl( \bar{I}_{k}^{(h)}\, \bar{I}_{k}^{(y)}\,
  \bar{I}_{k}^{(e)} \bigr)^{1/3}.
\label{eq:hdi}
\end{equation}
The uncertainty of the index is propagated by the delta method
\citep{casella2002,oehlert1992}. Taking the logarithm of Equation~\eqref{eq:hdi}
and differentiating, we obtain
\begin{equation}
\sigma_{\mathit{HDI}_{k}} = \frac{\mathit{HDI}_{k}}{3}
\sqrt{
  \left( \frac{\mathit{SE}_{k}^{(h)}}{\bar{I}_{k}^{(h)}} \right)^{2} +
  \left( \frac{\mathit{SE}_{k}^{(y)}}{\bar{I}_{k}^{(y)}} \right)^{2} +
  \left( \frac{\mathit{SE}_{k}^{(e)}}{\bar{I}_{k}^{(e)}} \right)^{2}
},
\label{eq:delta}
\end{equation}
and the 95\% confidence interval is
$\mathit{HDI}_{k} \pm 1.96\,\sigma_{\mathit{HDI}_{k}}$.
Equation~\eqref{eq:delta} assumes independence between the three dimensions, an
approximation discussed in Section~\ref{sec:discussion}: since there is no
feedback from the income and education dimensions onto health --- which is
anchored externally --- the residual correlation is attenuated, though not
eliminated, so that the resulting interval is slightly optimistic.

\subsection{Validation strategy}

The validation compares the reconstructed country-level HDI with the official
UNDP HDI for 2014, published in the Human Development Report 2015
\citep{undp2015}. Since the official index is computed from national accounts
and vital statistics, sources entirely independent of the survey microdata, the
agreement between the two constitutes a test of convergent validity. Three
metrics are adopted: the mean absolute error (MAE), the Pearson correlation,
which measures agreement in level, and the Spearman correlation, which assesses
the preservation of the ranking of countries. The whole implementation relies on
the NumPy \citep{harris2020}, SciPy \citep{virtanen2020} and pandas
\citep{mckinney2010} libraries.

\section{Results}
\label{sec:results}

After applying the validity criteria, converting missing-value codes into
missing data and excluding records without sufficient information, 39,706 of the
40,185 respondents remained, or 98.8\% of the sample, distributed across the 21
countries. The reconstruction of the index proceeded on this base, with
aggregation by country and by demographic breakdowns.

\subsection{Validation by country}

Table~\ref{tab:countries} presents, for each country, the three dimensional
indices, the reconstructed HDI, the official UNDP HDI for 2014 and the
difference between them. The agreement is substantial: the mean absolute error
between the reconstructed and the official index is 0.0099, the Pearson
correlation reaches 0.917, and the Spearman correlation, which gauges the
fidelity of the ranking, attains 0.849. Several countries lie less than 0.002
from the official value, among them Germany, France, Slovenia, Estonia, Poland
and Portugal, which indicates that the individual-level reconstruction not only
gets the mean level right, but also captures the relative position of the
countries.

\begin{table}[htbp]
\centering
\caption{Reconstructed HDI by country versus official UNDP HDI (2014).}
\label{tab:countries}
\small
\begin{tabular}{lrrrrrr}
\toprule
Country & Health & Income & Education & HDI est. & HDI UNDP & Difference \\
\midrule
Norway          & 0.955 & 0.943 & 0.918 & 0.939 & 0.944 & $-$0.005 \\
Netherlands     & 0.949 & 0.926 & 0.925 & 0.933 & 0.922 & $+$0.011 \\
Sweden          & 0.958 & 0.923 & 0.911 & 0.930 & 0.907 & $+$0.023 \\
Ireland         & 0.944 & 0.877 & 0.939 & 0.920 & 0.916 & $+$0.004 \\
Belgium         & 0.943 & 0.907 & 0.909 & 0.919 & 0.890 & $+$0.029 \\
Germany         & 0.940 & 0.915 & 0.892 & 0.915 & 0.916 & $-$0.001 \\
Finland         & 0.941 & 0.889 & 0.908 & 0.913 & 0.883 & $+$0.030 \\
United Kingdom  & 0.943 & 0.882 & 0.909 & 0.911 & 0.907 & $+$0.004 \\
Denmark         & 0.934 & 0.910 & 0.884 & 0.909 & 0.923 & $-$0.014 \\
Switzerland     & 0.972 & 0.942 & 0.808 & 0.905 & 0.930 & $-$0.025 \\
Austria         & 0.946 & 0.894 & 0.850 & 0.896 & 0.885 & $+$0.011 \\
Spain           & 0.973 & 0.847 & 0.872 & 0.896 & 0.876 & $+$0.020 \\
France          & 0.965 & 0.884 & 0.822 & 0.888 & 0.888 & $+$0.000 \\
Israel          & 0.956 & 0.856 & 0.852 & 0.887 & 0.894 & $-$0.007 \\
Slovenia        & 0.940 & 0.827 & 0.877 & 0.880 & 0.880 & $-$0.000 \\
Czechia         & 0.905 & 0.842 & 0.883 & 0.877 & 0.870 & $+$0.007 \\
Estonia         & 0.877 & 0.832 & 0.870 & 0.859 & 0.861 & $-$0.002 \\
Lithuania       & 0.839 & 0.829 & 0.866 & 0.845 & 0.839 & $+$0.006 \\
Poland          & 0.886 & 0.804 & 0.838 & 0.842 & 0.843 & $-$0.001 \\
Portugal        & 0.940 & 0.818 & 0.746 & 0.831 & 0.830 & $+$0.001 \\
Hungary         & 0.858 & 0.826 & 0.803 & 0.829 & 0.836 & $-$0.007 \\
\bottomrule
\end{tabular}
\end{table}

\begin{figure}[htbp]
\centering
\begin{subfigure}{\textwidth}
  \centering
  \includegraphics[width=\textwidth]{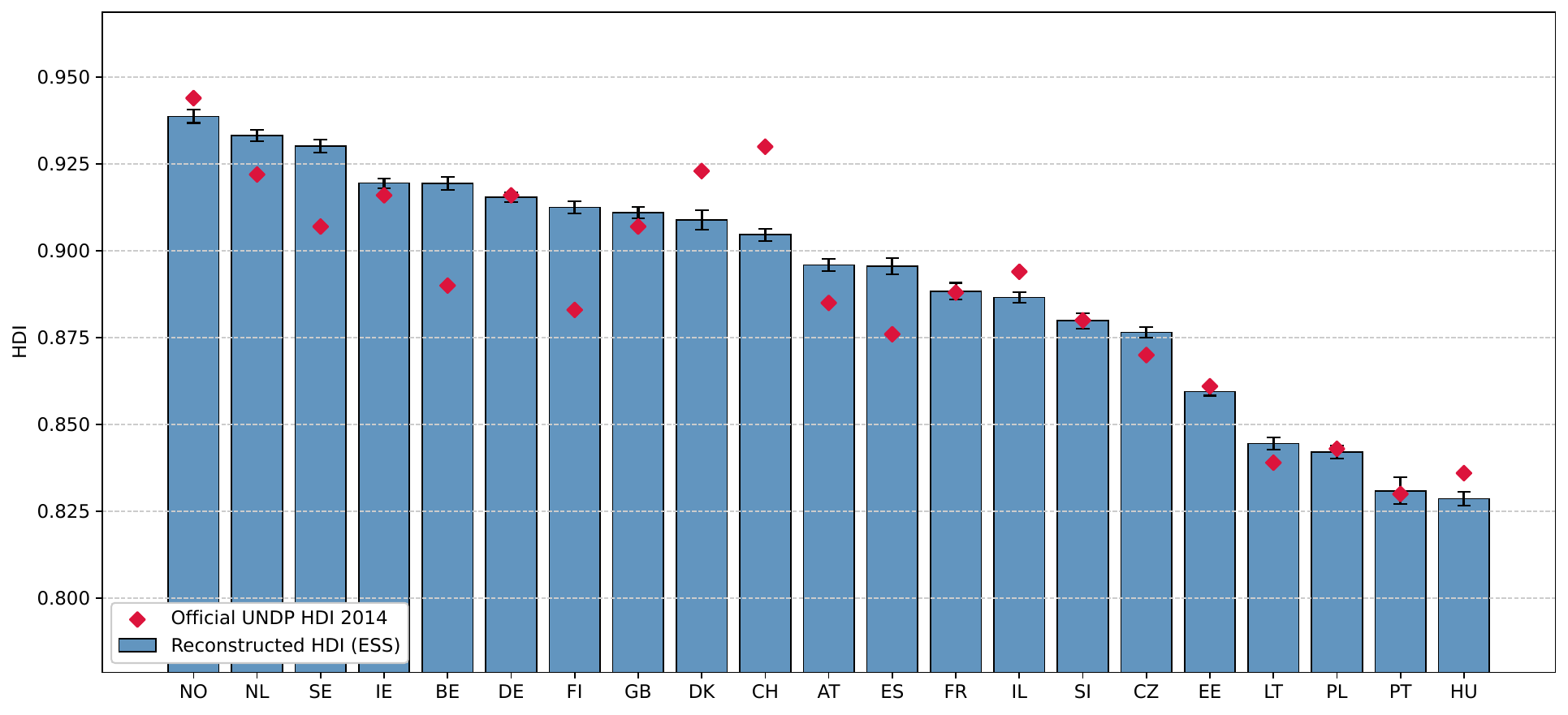}
  \caption{Reconstructed HDI versus official 2014 HDI.}
\end{subfigure}
\vspace{1em}
\begin{subfigure}{\textwidth}
  \centering
  \includegraphics[width=\textwidth]{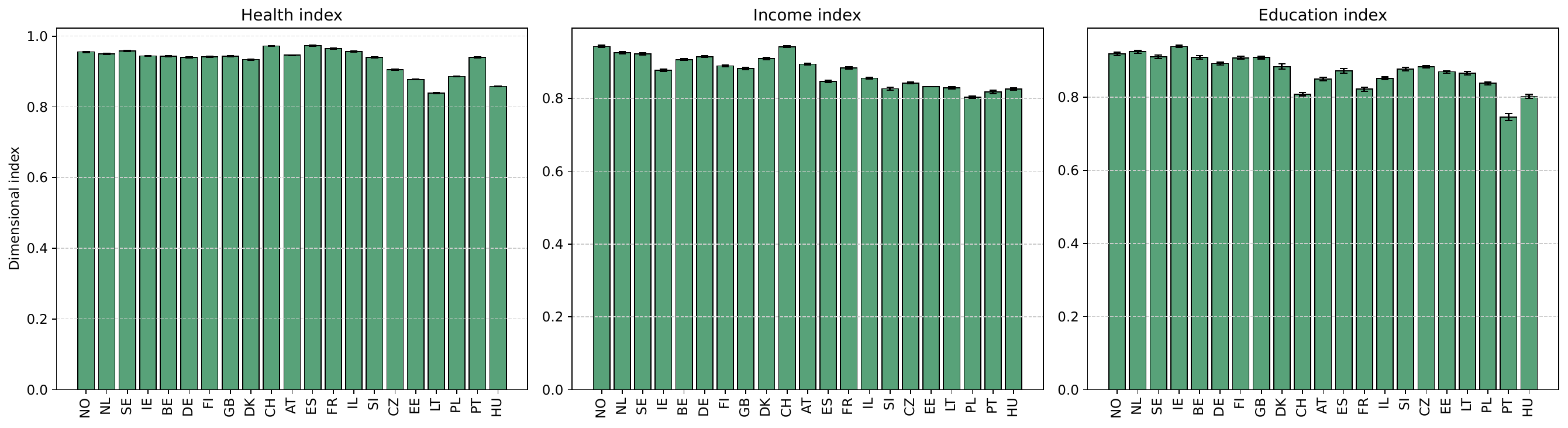}
  \caption{Contribution of the health, income and education dimensions to the
  reconstructed HDI.}
\end{subfigure}
\caption{Comparison and decomposition of the reconstructed HDI across the
countries analysed. Error bars denote 95\% confidence intervals.}
\label{fig:countries}
\end{figure}

Figure~\ref{fig:countries} summarises this comparison graphically, contrasting
the reconstructed HDI with the official one country by country
(Figure~\ref{fig:countries}a), and the decomposition of the index into its three
constituent dimensions (Figure~\ref{fig:countries}b), showing the contribution
of health, income and education to the aggregate value of each country.

\subsection{Breakdown by sex}

The HDI disaggregated by sex, considering all countries with sampling weights,
showed that men exhibit an HDI of 0.900 (95\% CI: 0.899--0.901), against 0.892
among women (95\% CI: 0.891--0.893), a modest difference whose 95\% confidence
intervals, narrow by virtue of the large sample size, do not overlap. The
differential is concentrated in the income (0.884 versus 0.872) and education
(0.874 versus 0.867) dimensions, whereas the health dimension remains close
between the sexes (0.943 among men and 0.938 among women).

\subsection{Breakdown by age band}

A clear age gradient is observed: the HDI rises from 0.900 in the 18--24 group
to a maximum of 0.914 in the 25--34 band, followed by a slight decline between
35 and 44 years (0.911) and 45 and 54 years (0.906), becoming more pronounced
between 55 and 64 years (0.890) and reaching 0.863 among individuals aged 65 or
over. This pattern reflects the interaction of two opposing forces across
dimensions --- younger cohorts benefit from greater schooling, resulting from
educational expansion, while the health dimension declines with age, as expected
from the modulation of life expectancy.

\subsection{Socioeconomic breakdowns: occupation and marital status}

The flexibility of the individual-level formulation manifests itself with
particular clarity in the socioeconomic breakdowns, inaccessible to the
aggregate approach. Table~\ref{tab:occupation} orders the HDI according to
declared occupational status, revealing a pronounced gradient: individuals in
paid work reach an HDI of 0.915, whereas the permanently sick or disabled stand
at 0.855, a difference of six hundredths, an order of magnitude comparable to
that separating the countries at the top and the bottom of
Table~\ref{tab:countries}.

\begin{table}[htbp]
\centering
\caption{HDI by occupational status (all countries, weighted).}
\label{tab:occupation}
\small
\begin{tabular}{lrrrrr}
\toprule
Occupational status & $n$ & Health & Income & Education & HDI \\
\midrule
Paid work                    & 19,912 & 0.947 & 0.898 & 0.900 & 0.915 \\
Education (student)          &  3,346 & 0.954 & 0.876 & 0.874 & 0.900 \\
Other                        &    478 & 0.942 & 0.874 & 0.862 & 0.892 \\
Housework, caring            &  2,666 & 0.942 & 0.864 & 0.852 & 0.885 \\
Unemployed                   &  2,123 & 0.946 & 0.829 & 0.864 & 0.878 \\
Retired                      & 10,000 & 0.925 & 0.857 & 0.815 & 0.865 \\
Permanently sick or disabled &  1,060 & 0.891 & 0.842 & 0.834 & 0.855 \\
\bottomrule
\end{tabular}
\end{table}

\begin{figure}[htbp]
\centering
\includegraphics[width=0.85\textwidth]{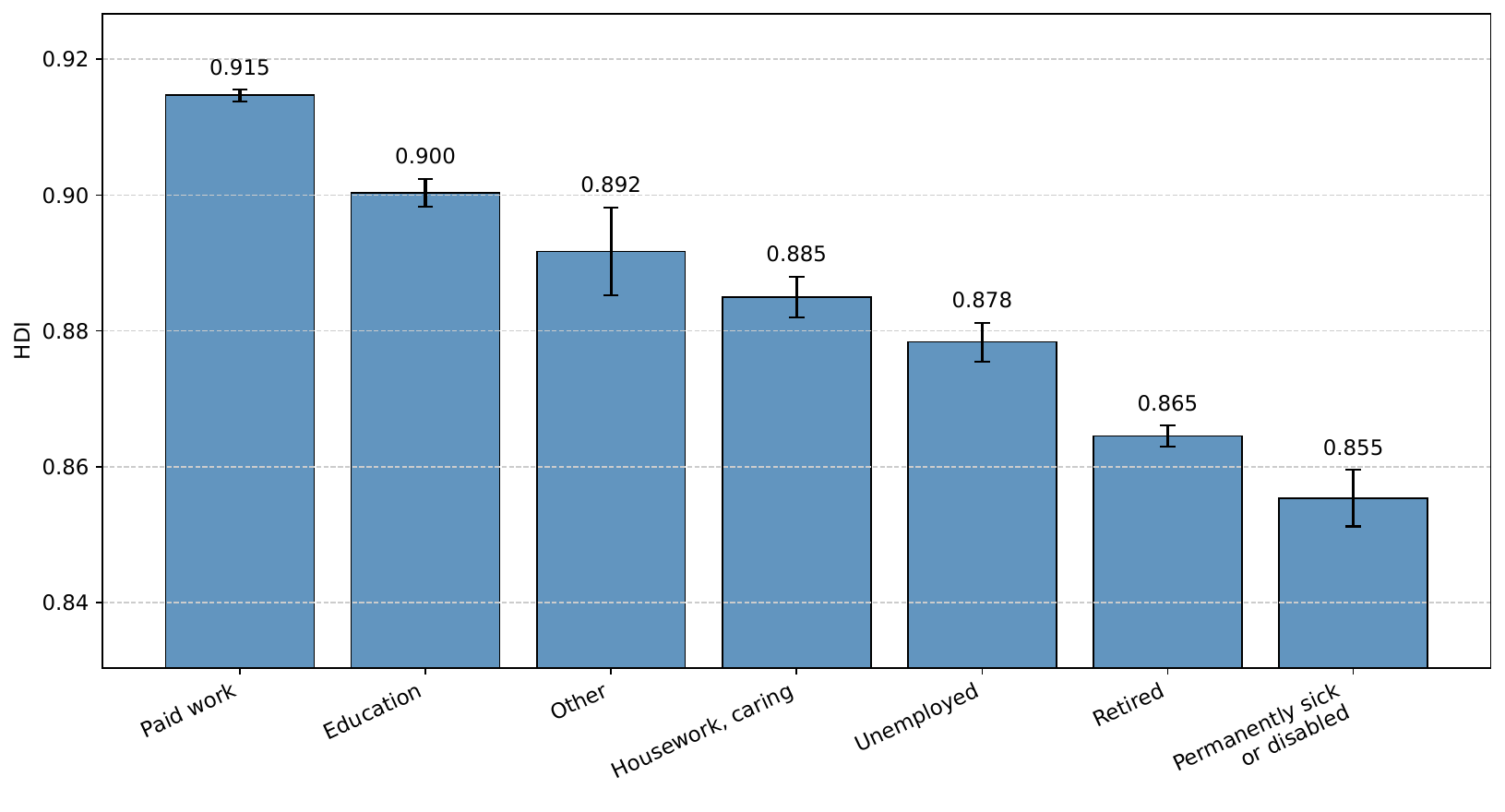}
\caption{HDI by occupational status, with 95\% confidence intervals.}
\label{fig:occupation}
\end{figure}

The decomposition is informative: the income dimension is lowest among the
unemployed (0.829), reflecting the loss of earnings associated with
unemployment; education is lowest among the retired (0.815), consistent with
older and less schooled cohorts; and health drops among the disabled (0.891), as
expected given the modulation by self-rated health. Figure~\ref{fig:occupation}
summarises this gradient with 95\% confidence intervals.

The breakdown by marital status exhibits a similar pattern. Never-married
individuals and those married or in a union present the highest HDIs (0.903 and
0.900, respectively), followed by the separated or divorced (0.885), while the
widowed record the lowest value (0.843). Among the widowed, the lowest values in
all three dimensions of the index are also observed, especially in education
(0.792) and income (0.825), while health presents a value of 0.916. This pattern
suggests that the differences observed in the HDI across groups are particularly
associated with the socioeconomic dimensions, and may also reflect differences
in the age composition of the groups.

\subsection{Territorial and origin breakdowns}

Breakdowns of a territorial and origin nature produce fainter gradients, a
finding that is relevant in itself. Table~\ref{tab:area} arranges the HDI by
type of area of residence: the difference between the suburbs of big cities
(0.912) and country villages (0.889) is only two hundredths, which suggests
that, in countries of very high human development, the classic urban--rural
divide is substantially attenuated. Curiously, the most isolated rural area
records the highest health score (0.948), even though its lower education score
depresses the aggregate index.

\begin{table}[htbp]
\centering
\caption{HDI by area of residence (all countries, weighted).}
\label{tab:area}
\small
\begin{tabular}{lrrrrr}
\toprule
Area of residence & $n$ & Health & Income & Education & HDI \\
\midrule
Suburbs or outskirts of big city & 4,682 & 0.947 & 0.896 & 0.892 & 0.912 \\
Farm or home in countryside      & 2,531 & 0.948 & 0.884 & 0.859 & 0.897 \\
Big city                         & 8,364 & 0.939 & 0.874 & 0.879 & 0.897 \\
Town or small city               & 12,759 & 0.940 & 0.876 & 0.874 & 0.896 \\
Country village                  & 11,293 & 0.939 & 0.875 & 0.856 & 0.889 \\
\bottomrule
\end{tabular}
\end{table}

The breakdowns by origin reinforce this reading. As to immigration status, those
born abroad exhibit a slightly higher HDI than natives (0.899 against 0.895), a
difference sustained above all by higher self-rated health (0.951 against
0.939), a pattern compatible with the so-called healthy migrant effect,
according to which migrant populations tend to present better initial health
than the destination population. Belonging to an ethnic minority, in turn,
barely differentiates the index (0.897 among minorities and 0.896 among
non-minorities), indicating the absence of an aggregate gap in this dimension
for the set of countries analysed.

\section{Discussion}
\label{sec:discussion}

The strong agreement between the reconstructed and the official HDI deserves
careful interpretation, since it is precisely there that the evidence of
validity resides. The official value is computed from national accounts, vital
statistics and educational records; the reconstructed value emerges from
individual responses to a survey, weighted and aggregated. These are independent
sources of information, subject to errors of a distinct nature. That two such
different methodological trajectories converge to values whose mean absolute
error is below 0.01 suggests that both capture the same underlying construct,
which constitutes the core of the convergent validity argument.

The decision to anchor the health dimension in official life expectancy, rather
than modelling it ecologically, proves decisive for this result. By treating
$\mathit{LE}_{p}$ as an anchor and restricting the individual contribution to a
centred modulation, Equation~(1) ensures that the reconstructed national mean
coincides with the official value by construction, while preserving an internal
individual ordering empirically justified by the association between self-rated
health and mortality \citep{idler1997}. This choice keeps the estimate immune to
the ecological fallacy \citep{robinson1950}, since it does not infer the
individual quantity from correlations observed between aggregates.

The main advantage of the individual-level methodology is its capacity to
generate demographic and socioeconomic breakdowns inaccessible to the aggregate
approach. The gradients by sex and by age are coherent with substantive
knowledge --- greater schooling in younger cohorts, decline of health with age
--- and illustrate the kind of diagnostic that only microdata enable.

The socioeconomic breakdowns go further and reveal gaps of a magnitude
comparable to the variation between countries: the distance of six hundredths
between the employed and the permanently disabled, or the 0.843 of the widowed
against the 0.903 of the never-married, evidence internal inequalities that the
national HDI, by construction, conceals. Such decompositions are not mere
descriptive curiosities: they identify which subpopulations concentrate
development deficits and in which dimension, information directly actionable by
public policy. It is also worth noting that territorial and origin breakdowns
produce faint gradients in this set of countries of very high development --- the
urban--rural divide is only two hundredths (Table~\ref{tab:area}) and
immigration status slightly favours those born abroad, in line with the healthy
migrant effect --- which suggests that, in these economies, the stratification of
human development operates more through occupational insertion and life cycle
than through territory or origin.

It is necessary, however, to acknowledge the limitations honestly. First, the
hypothesis of independence between dimensions in Equation~\eqref{eq:delta} makes
the confidence intervals slightly optimistic, since income, education and health
retain a residual correlation; a propagation incorporating the covariance
between dimensions would produce wider intervals. Second, the modulation of
health through the parameter $\beta$ is not equivalent to a subgroup-specific
life table, so that individual life expectancies should be read as plausible
displacements around the national mean, and not as direct measures of mortality
by stratum. Third, income is treated in a relative fashion, under a fixed Gini
coefficient of 0.30 common to all countries, which ignores the real
heterogeneity of inequality. Fourth, the coverage is limited to the 21 countries
of ESS7, all of very high human development, which restricts generalisation to
contexts of lower development.

The specific divergences are instructive. Switzerland is underestimated above
all in the education dimension, since the mean years of schooling declared in the
ESS lie below what the official indicators would suggest, depressing the
educational component and, consequently, the aggregate index. In the opposite
direction, Belgium, Finland and Sweden are overestimated, with positive
differences above 0.02, which may reflect both particularities of the sample
composition and effects of the health modulation and of the relative income
structure. Such discrepancies, far from invalidating the approach, delimit its
domain of applicability and point to paths for refinement.

\section{Conclusion and Future Work}
\label{sec:conclusion}

This work proposed a methodology for reconstructing the Human Development Index
from individual-level survey microdata and validated it against the official
UNDP index, demonstrating that the HDI reconstructed from the 39,706 valid
respondents of the European Social Survey, Round 7, reproduces the official 2014
HDI with a mean absolute error of 0.0099 and a Pearson correlation of 0.917. The
central methodological choices --- anchoring the health dimension in national
life expectancy, using actual sampling standard errors, and treating income
through a single route coherent with the UNDP standard --- proved decisive for
the observed agreement. In answer to the research questions, we conclude that it
is indeed possible to reconstruct, from microdata, a country-level HDI that
adheres to the official one, and that the individual-level formulation enables
demographic disaggregations with quantification of uncertainty.

Several developments deserve future investigation. Calibrating the Gini
coefficient by country, replacing the fixed value, would bring the income
structure closer to the real inequality of each nation. Incorporating
subgroup-specific life tables would give the health dimension a firmer
demographic foundation than the modulation by $\beta$. Uncertainty propagation
could be improved by including the covariance between dimensions, producing more
conservative and realistic confidence intervals. Inequality-adjusted variants of
the index, sensitive to the internal distribution of each dimension, are
naturally enabled by access to microdata. Finally, applying the methodology to
Round 11 of the European Social Survey, referring to 2023, would allow the
temporal evolution of disaggregated human development to be analysed, opening
the way to longitudinal studies.

\section*{Data and Code Availability}

The ESS7 microdata are distributed by Sikt/NSD under
DOI~\href{https://doi.org/10.21338/ess7e02_3}{10.21338/ess7e02\_3}; the World
Development Indicators series are openly available from the World Bank. The
analysis notebook that reproduces every table, figure and reported value of this
article --- together with the processing scripts and the open national inputs ---
is archived at Zenodo under
DOI~\href{https://doi.org/10.5281/zenodo.22714205}{10.5281/zenodo.22714205}
\citep{schmitz2026notebook}.

\bibliography{references}

\end{document}